\documentclass[nonacm]{acmart}

\begin{document}

\acmBooktitle{arXiv}

\title{Cybersecurity competence of older adult users of mobile devices}

\author{Simon Vrhovec}
\email{simon.vrhovec@um.si}
\affiliation{
  \institution{University of Maribor}
  \city{Maribor}
  \country{Slovenia}
}
\author{Igor Bernik}
\email{igor.bernik@um.si}
\affiliation{
  \institution{University of Maribor}
  \city{Maribor}
  \country{Slovenia}
}
\author{Damjan Fujs}
\email{damjan.fujs@fri.uni-lj.si}
\affiliation{
  \institution{University of Ljubljana}
  \city{Ljubljana}
  \country{Slovenia}
}
\author{Damjan Vavpotič}
\email{damjan.vavpotic@fri.uni-lj.si}
\affiliation{
  \institution{University of Ljubljana}
  \city{Ljubljana}
  \country{Slovenia}
}
\renewcommand{\shortauthors}{Vrhovec et al.}

\begin{abstract}
This work reports on a cross-sectional study on device proficiency, support availability and cybersecurity competence of older adult users of smartphones and/or tablets. Results indicate that cybersecurity competence is associated with both device proficiency and support availability although the variance explained is relatively low. There were no differences in cybersecurity competence between users and non-users of either mobile devices. Users of both smartphones and tablets had significantly higher device proficiency than non-users. Users of tablets had significantly higher support availability than non-users while there were no significant differences between users and non-users of smartphones.
\end{abstract}


\maketitle

\section{Introduction}

Smartphones and tablets are among the most used smart devices by older adults \cite{Mihelic2022}. These mobile devices enable them to keep contact with their families, read news, make video calls, share photos, and even shop online. Adults over 60 are continually among age groups most victimized by cyber crimes with losses far surpassing the losses of other age groups \cite{FBI2023}. Recent studies identified some key dimensions, such as security awareness, support availability and digital literacy, which are likely to influence older adult cybersecurity performance \cite{Morrison2023}. It however remains unclear how these factors are related to each other, and whether there are differences between users and non-users of certain types of devices. This work aims to investigate whether device proficiency (a dimension of digital literacy) and support availability are associated with cybersecurity competence (enveloping awareness) of older adults, and explore differences in these factors between users and non-users of mobile devices.

\begin{table*}[ht!]
\caption{\label{table:t-tests} Differences between users and non-users of mobile devices were investigated with independent samples $t$-tests.}
\footnotesize
\begin{tabular}{llrrrr}
\toprule
Device & Construct & Users & Non-users & $t$ & $p$ \\
\midrule
Smartphone &  & $n=736$ & $n=45$ &  &  \\
 & Cybersecurity competence & $4.28$ & $4.23$ & $-0.667$ & $0.505$ \\
 & Device proficiency & $4.27$ & $3.45$ & $-5.172^{***}$ & $<0.001$ \\
 & Support availability & $3.75$ & $3.29$ & $-1.896$ & $0.064$ \\
Tablet &  & $n=488$ & $n=293$ &  &  \\
 & Cybersecurity competence & $4.31$ & $4.24$ & $-1.787$ & $0.074$ \\
 & Device proficiency & $4.33$ & $4.05$ & $-5.094^{***}$ & $<0.001$ \\
 & Support availability & $3.79$ & $3.58$ & $-2.245^{*}$ & $0.025$ \\
\bottomrule
\end{tabular}
\\
Notes: $^{***}p<0.001$, $^{*}p<0.05$
\end{table*}

\section{Research methodology}

This study employed a cross-sectional research design. We conducted a survey of older adults aged 65 or more residing in the UK through the \textit{Prolific} platform between 17 and 18 January 2024. The survey questionnaire included previously validated items adapted to the context of our study. It measured \textit{support availability} \cite{Ma2016} (three items) and \textit{knowledge} \cite{Chen2021} (three items) which are two dimensions of facilitating conditions \cite{Mihelic2023}, \textit{mobile device proficiency} with MDPQ (16 items) \cite{Roque2018} and \textit{cybersecurity competence} with S-HAIS-Q (39 items) \cite{Hadlington2020}. All items were measured with a 5-point Likert scale from 1 to 5. The study proposal was approved by the Research Ethics Committee of the University of Maribor, Faculty of Criminal Justice and Security [2711-2023].

We received a total of 804 responses. After excluding responses with wrong Prolific ID (13), responses with over 10 percent missing values (7) and respondents who were less than 65 years old (2), we were left with $N=782$ useful responses. $55.8\%$ of respondents were female, $44.1\%$ were male. Formal education of respondents ranged from finished high school or less ($47.7\%$) and received Bachelor's degree ($37.9\%$) to received Master's degree ($11.4\%$) and received PhD degree ($3.1\%$). While $77.2\%$ respondents were retired, $22.8\%$ were still employed or self-employed. $60.5\%$ of respondents lived in an urban, and $39.5\%$ lived in a rural environment.

The reliability and validity of the survey instrument was assessed with exploratory factor analysis (EFA) and Cronbach's alpha (CA). First, we excluded 25 \textit{cybersecurity competence} items and four \textit{mobile device proficiency} items that did not follow a normal distribution (i.e., $|skewness| > 2$ or $|kurtosis| > 7$). Next, we extracted factors with EFA by using the \textit{principal axis factoring} method and \textit{promax} rotation. After excluding two additional \textit{cybersecurity competence} items due to low factor loading, we found an acceptable solution with three underlying factors. The first factor included the remaining 12 items for \textit{cybersecurity competence} ($CA=0.795$), the second factor included all three items for \textit{support availability} ($CA=0.967$), and the third factor included 12 items for \textit{mobile device proficiency} and three items for \textit{knowledge} ($CA=0.921$). The third factor indicates that the two constructs are very close in terms of their measurement. This is consistent with theory since self-reported knowledge necessary to use mobile devices can be directly related to proficiency of using mobile devices. To avoid confusion, the third (merged) factor was renamed to \textit{device proficiency}. CA of all three final constructs was above the 0.70 threshold for acceptable reliability.

\section{Cybersecurity competence}

To investigate the determinants of cybersecurity competence, we performed a linear regression of \textit{device proficiency} and \textit{support availability} (independent variables) on \textit{cybersecurity competence} (dependent variable). These results support the hypothesis that both device proficiency ($\beta = 0.130$, $p<0.001$) and support availability ($\beta = 0.072$, $p<0.001$) are directly related to cybersecurity competence. A relatively low variance explained ($R^2 = 0.082$) however suggests the existence of other significant factors not accounted for in our study, such as stress, financial health and perceived responsibility \cite{Morrison2023}.

\section{Users vs. non-users of mobile devices}

We used independent samples $t$-tests to explore differences between users and non-users of mobile devices since all three aggregated variables followed a normal distribution. The results of these tests are presented in Table~\ref{table:t-tests}. First, \textit{cybersecurity competence} does not seem to differ between users and non-users of neither mobile devices. These results indicate that cybersecurity competence may be generic, likely acquired with experience with any device. Second, the most notable difference between users and non-users of both mobile devices was with \textit{device proficiency} ($p<0.001$). These results confirm the assumption that older adults who use a device become more proficient with it. It is also likely that older adults who are more proficient with a specific device are more likely to use it. Third, there were no significant differences in \textit{support availability} between users and non-users of smartphones. There were however significant differences between users and non-users of tablets ($p<0.05$). These results point to the need for support in adoption of tablets but not smartphones. These significant differences for \textit{device proficiency} and \textit{support availability} coupled with the lack of a significant difference for \textit{cybersecurity competence} between users and non-users of specific mobile devices appear to further confirm the generic nature of cybersecurity competence.

\section{Limitations and future work}

Although this study provided some useful insights, there are some limitations that the readers should note. First, respondents were recruited through the \textit{Prolific} platform. Studies through other channels reaching a more general population would be therefore needed to investigate whether the results of this study are generalizable to the general population of older adults. Second, this study indicated notable issues with the measurement of \textit{cybersecurity competence} with S-HAIS-Q as 27 out of 39 items were excluded due to poor measurement performance. The issues with non-normality may be partially attributed to the sample (22 excluded items had their mean above $4.6$ on a 5-point scale). The development of alternative measures of cybersecurity competence would be thus beneficial to measure cybersecurity competence of well-performing populations. Third, the merging of constructs \textit{mobile device proficiency} and \textit{knowledge} into a single factor indicates some potential measurement optimization for device proficiency. Future studies would however be needed to determine whether this indication applies beyond the studied population.

\begin{acks}
\small
This work was partially funded by the Slovenian Research Agency [grant number J5-3111].
\end{acks}

\bibliographystyle{ACM-Reference-Format}


\end{document}